\documentclass[aps,pra,twocolumn,showpacs]{revtex4-1}
\usepackage{graphicx}
\usepackage{amsmath}
\usepackage{bm}
\newcommand{\beq}{\begin{equation}}
\newcommand{\eeq}{\end{equation}}
\newcommand{\beqa}{\begin{eqnarray}}
\newcommand{\eeqa}{\end{eqnarray}}

\usepackage{colordvi}
\usepackage{color}

\begin{document}

\title{Timescales of tunneling decay of a localized state}

\author{Yue Ban$^{1}$, E. Ya. Sherman$^{1,2}$, J. G. Muga$^{1}$, and M. B\"{u}ttiker$^{3}$}
\affiliation{$^1$Department of Physical Chemistry, Universidad del Pa\'is Vasco UPV-EHU,
48080 Bilbao, Spain \\
$^2$ IKERBASQUE, Basque Foundation for Science, 48011, Bilbao, Spain\\
$^3$D\'epartement de Physique Th\'eorique, Universit\'e de Gen\'eve, CH-1211 Gen\'eve 4, Switzerland}

\begin{abstract}
Motivated by recent time domain experiments on ultrafast atom ionization,
we analyze the transients and timescales
that characterize, besides the relatively long lifetime, the decay by tunneling of a
localized  state. While the tunneling starts immediately, some time is
required for the outgoing flux to develop. This short-term behavior depends strongly
on the initial state. For the initial state tightly localized so that the initial transients
are dominated by over-the-barrier motion, the timescale for the flux propagation through the barrier is close to
the B\"uttiker-Landauer traversal time. Then a quasistationary, slowly decay process
follows, which sets ideal conditions for observing diffraction in time
at longer times and distances. To define operationally a tunnelling time at the barrier edge,
we extrapolate backwards the propagation of the wave packet escaped from the potential.
This extrapolated time is considerably longer than the timescale of the flux and density buildup
at the barrier edge.
\end{abstract}


\maketitle

\section{Introduction}
Tunneling, being one of the most fundamental concepts in quantum mechanics,
remains a source of strong theoretical and experimental
controversies on the relevant timescales of the process \cite{MacColl,Mugabook}.
A typical tunneling setting is the scattering process, where a wavepacket is reflected from and
transmitted through a barrier higher than the incident particle
energy. In this case, the group delay  \cite{Wigner}, defined in
terms of the energy derivatives of the reflection or transmission
phase shift, describes the motion of the wave packet peak. It was also
found that the group delay for particles tunneling through an opaque
barrier is independent of the barrier width (the ``Hartman effect''  \cite{Hartman}).

The measurement of the time spent by a tunneling particle in the classically forbidden
region can be based on the approach of Baz and Rybachenko
\cite{Baz,Rybachenko}, which uses the Larmor precession of a particle
with a magnetic moment in a
weak magnetic field or in effective fields in solids due
to the spin-orbit coupling \cite{Sherman10,Appelbaum10}.
In classically forbidden regions there is not only precession but also a rotation  \cite{dwell}
of the moment into the direction of the field.
B\"{u}ttiker and Landauer \cite{Buttiker-Landauer} analyzed tunneling
through an oscillating rectangular
barrier and found an interaction
time which is closely related to the rotation of the magnetic
moment in a magnetic field \cite{Buttiker-Landauer,dwell}.
A tunneling time has been measured for electromagnetic waves
passing through inhomogeneous optical structures \cite{Chiao-optical barrier,Chiao-photonic band}
and waveguides \cite{Stein,Mugn}. Interestingly, in graphene, a single layer of carbon atoms,
which provides another example of massless particles, the transport occurs via
evanescent waves, and the Wigner-Smith delay is
{\it linear} in the tunneling distance \cite{Sepkhanov09}.

Brouard, Sala, and Muga, using scattering theory
projectors for to-be-transmitted/reflected wavepacket parts and
for localizing the particle
at the barrier, set a formal framework that unified many of the existing proposals
pointing out that the multiple time scales correspond to different
quantizations of the classical transmission time,
due to the noncommuting observables and possible orderings involved  \cite{BSM}.
This clarified the meaning of different partitions of the dwell time into transmitted, reflected and interference components.
Another research line has been the
investigation of characteristic times for the transient dynamics of the wave function (e.g. the forerunners) in a plethora of potential configurations and initial conditions using asymptotic methods  \cite{BM,Buttiker_Thomas,transient}.
For specially prepared states, in particular for confined or semiconfined initial waves with a flat density, these transients show diffraction in time \cite{DIT,transient}, i.e. temporal oscillations
reminiscent of spatial Fresnel-diffraction by a sharp edge  \cite{DIT}.
Also, the analysis of the partial density of states \cite{density of states} provides
another approach to  the understanding of the tunneling process.

An important tunneling-dependent phenomenon
is the decay of a metastable system  \cite{Decay1,Decay2,Kalbermann,Kelkar-EPL}, related, e.g.,
to the state ionization in optics and to the
discharging of a capacitor in mesoscopic physics \cite{Buttiker93,Gabelli,Nigg06,Fev2007,Moskalets08,Keeling08,Buttiker10}.
Compared to the full scattering problem, the decay configuration, or
``half-scattering'' problem, has been frequently considered unproblematic
because of the absence of the transmitted/reflected wavepacket splitting and the
existence of a well known characterization in terms of resonance lifetimes.
In fact, one may still pose classically minded questions on the tunnelling time similar to the ones
in the scattering configuration, however, without obvious
answers. A particle may wander in the trapping well for a while and then escape
through the barrier. For such a history the lifetime would be a waiting
time in the well plus a tunneling time.
Can this quantity be defined and measured in a sensible way?
The understanding can be based on the analysis of quantum
interference among the Feynman paths  \cite{Dmitri} or on the
consideration of the quantities defined by operational procedures, as presented
in this paper.

In recent years, the
techniques of atom ionization by a strong laser field and attosecond
probing of electron dynamics opened a new venue for experimental
studies of the tunneling times.
The measurement of ionization of He atoms \cite{science} holds the promise to observe
the tunneling delay of electrons in real time. In the experiment access to
the dynamics at the tunneling time scale is gained through extrapolation of
long time measurement to shorter times by assuming that a particle that has
escaped through an energetically forbidden into a classical allowed
region, follows classical dynamical laws. Using these laws and scattering
data the moment of escape from the tunneling region into the classical
region is determined. A related experiment \cite{NRC} measures the
perpendicular distribution of the nascent quantum mechanical momentum uncertainty
of the initial state as it is revealed by tunneling.

The aim of this paper is to study the dynamics of the tunneling-induced
decay of a localized state and to investigate to what extent we can extract
information on the short time behavior from the long time dynamics.
We introduce a simple quantum mechanical model,
which simulates the experimental measurement of
tunneling time and allows for opening and closing the barrier \cite{science}.
We calculate the probability density and flux to
study the short-term dynamics near the barrier
and the long-term dynamics far away from it.
For an opaque barrier we observe a relatively short transient time scale, where the tunneling is developed
into a quasi-steady decay process. The details of the initial state have an important impact on the
transients. At large distances, the flux and the density
start to grow at long times, reach a peak and then decrease by
diffraction-in-time oscillations.
We found that the tunneling time obtained by extrapolating
the particle motion from the position of the remote detector to the
right edge of the barrier is not directly related to the timescale
of the outgoing flux buildup.
\section{General description of tunneling and model potential}
To study the tunneling dynamics, we consider the potential $U(x,t)$,
infinite at $x< 0$.
At $t<0$ the potential holds bound states with wavefunction $\varphi_j(x)$ and energy $E_j$;
it changes at $t=0$ to allow the tunneling, and changes back at $t=t_0$ to its initial form.
The entire time dependence is
\begin{eqnarray} \ U\left(x,t\right)= \left\{
\begin{array}{ll}
U_1(x)  &~~ (t<0 ~~\mbox{and}~~ t>t_0){,}
\\
U_2(x) &~~~ (0<t<t_0){.}
\end{array}
\right.
\end{eqnarray}
The initial state $\Psi(x,t=0)$ prepared at $t=0$
begins to evolve at $t>0$ and the probability to find the
electron inside the potential decreases. This is similar to the ionization of an atom by
a strong laser field, which lets a valence electron to tunnel
through the barrier. At the closing time
$t_0$, the potential becomes $U_1(x)$ again and the decay
terminates.
In the time interval $0<t<t_0$ the wave function $\Psi (x,t)$ can be expressed as
\begin{eqnarray}
\label{wave function1}
\Psi(x,0<t<t_0)= \int_{0}^\infty G (k) \phi_k
(x) \exp \left(-\frac{i k^2 t}{2}\right) dk,
\end{eqnarray}
with
\begin{eqnarray}
\label{coefficient1}G(k) =\int_0^\infty \Psi (x,0)\phi_k(x) dx,
\end{eqnarray}
where $\phi_k(x)$ are delocalized real states in the potential $U_2(x)$ with the energy $E=k^2/2$.
We use here $\hbar\equiv m\equiv 1$ units, where $m$ is the electron mass.

\begin{figure}[ht]
\begin{center}
\scalebox{0.6}[0.65]{\includegraphics{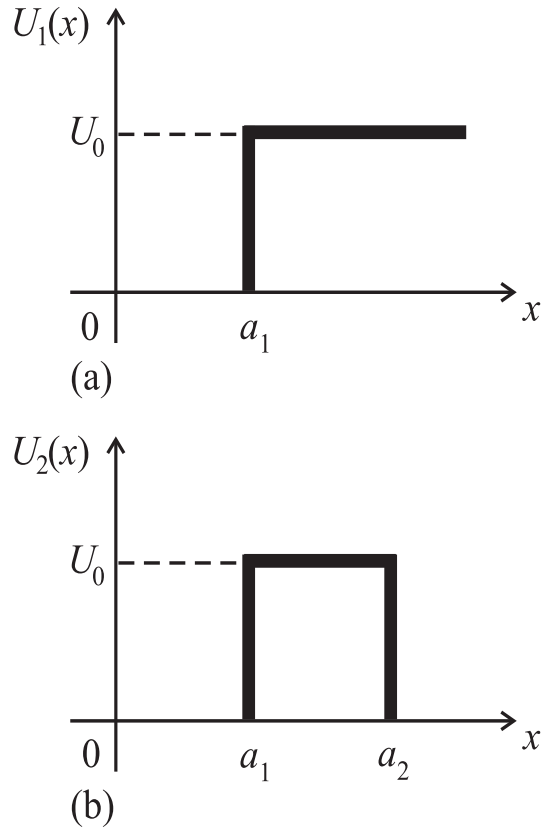}}
\caption{The time-dependent potential $U_1(x)$ at $t=0$ and $t>t_0$ and $U_2(x)$ within the interval $0<t<t_0$.
(a) shows that $U_1(x)$ is a step in the positive-half plane, which is $0$ from $x=0$ to
$x=a_1$ and $U_0$ from $x=a_1$ to infinity, while (b) presents a barrier which extends from $x=a_1$ to $x=a_2$ and is of height of $U_0$.
The potential is always infinite in the negative halfplane.}
\label{model}
\end{center}
\end{figure}

After closing the barrier at $t=t_0$, the potential
gets its original form, and the wave function can
be expressed as
\begin{eqnarray}
\label{wave function21}
\Psi(x,t>t_0) &=&
\sum_{j}B_{j}\varphi_j(x)\exp\left[-iE_{j}(t-t_0)\right]
\\ \nonumber &+&
\int_{\sqrt{2 U_0}}^\infty
B(k) \varphi_k(x) \exp \left[-\frac{i k^2 (t-t_0)}{2}\right] dk,
\end{eqnarray}
where summation is extended over the bound states of the initial potential, and
\begin{eqnarray}
&&\label{coefficientB0}B_{j}= \int_0^\infty \Psi(x,t_0)
\varphi_{j}(x)dx, \\
&&\label{coefficientB}B(k)= \int_0^\infty \Psi(x,t_0) \varphi_k(x)dx,
\end{eqnarray}
with $\varphi_k(x)$ being the continuum states in the potential $U_1(x)$.
With Eqs. (\ref{wave function1}) and (\ref{wave function21}) we have fully
specified the time evolution of the escape problem.

\begin{figure}[ht] \begin{center}
 \scalebox{0.8}[0.8]{
\includegraphics*{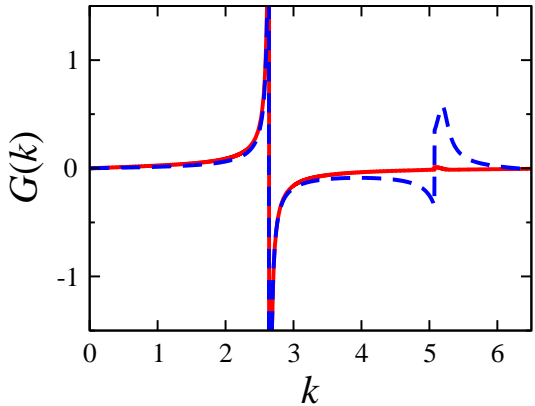}}
 \caption{
(Color online)~ The matrix element $G(k)$ of the initial state with the delocalized
states of the potential $U_2$. $k$ is the wave vector of the
delocalized states. The solid line corresponds to the initial state $\Psi(x,0)=\varphi_0(x),$
which is a bound state of $U_1$. For this case the barrier in $U_2$ extends from
$a_1 = 1$ to $a_2 = 1.4$. The dashed line corresponds to a sinusoidal
initial state  $\Psi(x,0)=\sqrt{2} \sin(\pi x)$ and to a  barrier in $U_2$ extending from $a_1 =1$ to
$a_2 = 1.42$. In both cases, $\kappa d=2$ and $U_0=16$.}
\label{Gk} \end{center} \end{figure}

To specify the model, we assume that at $t<0$
the potential $U_1(x)$ is a step in the
positive half-axis, being zero at $0<x<a_1$ and $U_0$
at $x>a_1$, as shown in Fig.\ref{model}(a).
At $t=0$ the potential changes to $U_2(x),$ which is a rectangular barrier
of the height $U_0$ extended from $a_1$ to $a_2$, as shown in
Fig. \ref{model}(b). The rectangular barrier being the simplest example
of an everywhere finite potential, where the wavefunction and its
derivative are continuous, allows for the exact analysis
of the dynamics. We put here $a_1\equiv1$ and
measure the time, energy, and momentum in the corresponding units.

The delocalized eigenstates of the Hamiltonian corresponding to $U_1(x)$ have the form:
\begin{equation}
\label{eigenstateinwell} \varphi_k(x)= \left\{
\begin{array}{ll}
C_1(k) \sin(k x),  &~~ (0<x<a_1) \\
\sqrt{{2}/{\pi}} \sin[qx+\theta_{1}(k)], &~~ (x>a_1)
\end{array}
\right.
\end{equation}
where $q=\sqrt{k^2-2U_{0}}$, $k>\sqrt{2U_{0}}$ with $C_1(k)$ and $\theta_{1}(k)$
determined by the boundary conditions at $x=a_{1}$, and the norm
is determined by $\langle\varphi_{k'}|\varphi_{k}\rangle=\delta(q-q')$.

The eigenstate of the Hamiltonian corresponding to $U_2(x)$ is
\begin{eqnarray}
\label{eigenstateinbarrier}\ \phi_k(x)= \left\{
\begin{array}{ll}
C(k) \sin(k x),  &(0<x<a_1) \\
D(k) e^{-\kappa_{k}x} + F(k) e^{\kappa_{k}x}, &(a_1<x<a_2) \\
\sqrt{{2}/{\pi}} \sin[k x+\theta(k)], &(x>a_2)
\end{array}
\right.
\end{eqnarray}
with the normalization condition $\langle\phi_{k'}|\phi_{k}\rangle=\delta(k-k')$.
The coefficients
$C(k)$, $D(k)$, $F(k)$ and the phase $\theta(k)$ satisfy the
boundary conditions of the potential $U_2(x)$, and $\kappa_{k}=\sqrt{2U_0-k^2}$.
In the tunneling regime $k<\sqrt{2U_0}$, while in the propagating regime $k>\sqrt{2 U_0}$, and $i\kappa_{k}$
is substituted by $q$ defined below Eq. (\ref{eigenstateinwell}).

\section{Tunneling dynamics at short and long times.}

\subsection{Decay of the state}
The decay rate is described by the probability to find the
particle in the well from $x=0$ to $x=a_1$ defined as
\begin{eqnarray}
\label{probability1} w_1(t)=\int_0^{a_1} \Psi^* (x,t) \Psi (x,t) dx.
\end{eqnarray}
The lifetime $t_l$, which also can be seen as the dwell
time, an important time scale introduced to
characterize the decay rate, is the time that it takes for the
relative probability in the well $w_1(t)/w_1(0)$ to decay to $1/e$.
For an opaque barrier the decay time can be written approximately as
$t_l\approx AT_{k_0} \exp (2 \kappa d)$, with $\kappa=\sqrt{2 (U_0-E_0)}$ and 
$d=a_2-a_1$, where $k_0=\sqrt{2E_{0}}$,
$T_{k_0}=2/k_0$ is the period of motion for a particle in the well,
$\exp(-2\kappa d)\ll1$ is the tunneling rate, and
$A$ is a barrier-dependent coefficient of order $1$.
Therefore, $\kappa d$ is a significant physical parameter with
regard to the decay process and the tunneling time. Analysis of different
examples where this simple expression for the lifetime
does hold can be found in Ref. \cite{Kelkar-EPL}.

\begin{figure}[]
\begin{center}
\scalebox{0.8}[0.8]{\includegraphics*{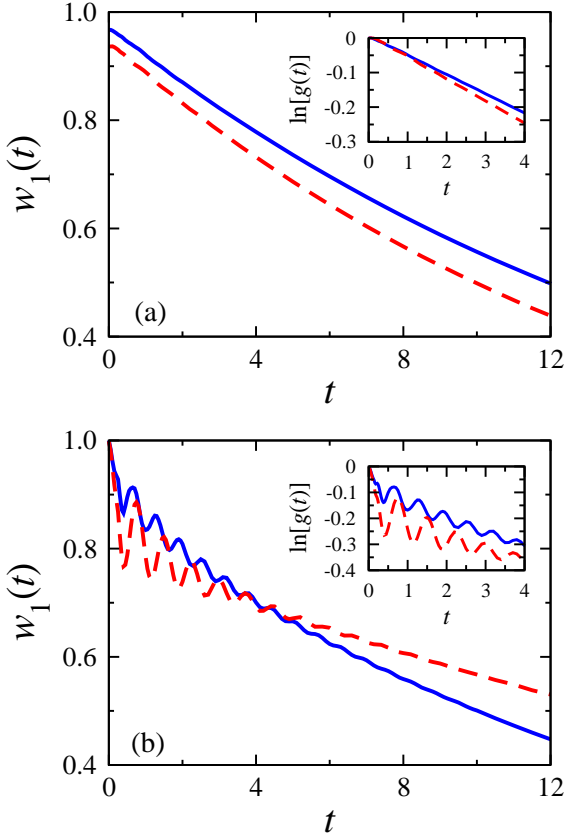}}
\caption{ (Color online)~ The probability to find  the particle in the well from $0$
 to $a_1$. (a) The solid and the dashed lines are for two initial states
$\Psi(x,0)=\varphi_0(x)$, one for $U_0=16$
and the other for $U_0=10$. (b)
 The solid and the dashed lines represent the barriers  $U_0=16$,
and $U_0=10$ with the same initial state $\Psi(x,0)=\sqrt{2} \sin(\pi
 x)$. The insets exhibit $\ln[g(t)]$
 at small time scale, where $g(t)\equiv w_1(t)/w_1(0)$. For all curves $\kappa d=2$.}
\label{Probability} \end{center} \end{figure}

To study the dynamics, we first set the initial state as the ground state of the Hamiltonian
corresponding to $U_1(x)$ for a typical potential.
For example, for $U_0=16$  there are two bound
eigenstates: the ground $\varphi_0(x)$ (energy $E_0$) and the excited state $\varphi_1(x)$ (energy $E_1$).
The energy $E_0$ for $\Psi(x,0)=\varphi_0(x)$ is $3.52$. By choosing different $a_2$,
one can modify the transparency of the barrier.
For example, if $\kappa d=2$, corresponding to $a_2=1.4,$ this barrier is a moderately
opaque one. The corresponding $G(k)$ is shown in Fig. \ref{Gk}. Under the conditions of $\Psi(x,0)=\varphi_0(x)$ the lifetime
in this potential is $t_l=17.5$, as illustrated in Fig. \ref{Probability}(a).
Calculated real and imaginary parts of $\Psi(x,t)$ after
some short transients show fast oscillations with the envelope
function $\exp\left(-t/2t_{l}\right)$. This behavior implies that
in terms of the poles in the complex energy plane, the states we consider
correspond to the simple Breit-Wigner resonances. Detailed analysis
of  various types of resonances and their relation to different timescales
can be found in Refs. \cite{Kelkar1,Kelkar2}.
To clarify the influence of the initial
state on tunneling, we alter it into the ground state of an
infinite-wall potential, $\Psi(x,0)=\sqrt{2}\sin(\pi x)$. Therefore, $E_0$ is $\pi^{2}/2$
and $a_2$ becomes $1.42$ to keep $\kappa d$ unchanged.
As shown in Fig. \ref{Gk}, the coefficient
$G(k)$ in Eq. (\ref{coefficient1}) has two contributions, related to the resonances
corresponding to the bound states of the
initial potential. For $\Psi(x,0)=\varphi_0(x)$, the second
one, corresponding to the first excited state with a fast decay, is extremely weak.
The presence of more than one bound state in the initial potential
combined with the condition of low transparency of the barrier leads to important consequences
for the short-time scale tunneling dynamics.

Although all the probabilities decrease exponentially at long times,
showing the general feature of the process,
the ones with $\Psi(x,0)=\sqrt{2} \sin(\pi x)$ in Fig.
\ref{Probability}(b) oscillate fast,
while those with $\Psi(x,0)=\varphi_0(x)$ in (a) decay smoothly.
This is because $\Psi(x,0)=\sqrt{2} \sin(\pi x)$ has
a significant contribution of the second anomaly shown in Fig.
\ref{Gk} leading to interference with the ``ground state resonance".
\subsection{Short-term dynamics}
In this subsection we address the beginning of the tunneling by studying
the short-time dynamics of the flux
\begin{eqnarray}
\label{flux definition}  J(x,t)=\frac{1}{2 i} \left[\Psi^*(x,t)
\frac{\partial \Psi(x,t)}{\partial x}-\frac{\partial
\Psi^*(x,t)}{\partial x} \Psi(x,t)\right],
\end{eqnarray}
and the density,
\begin{eqnarray}
\label{density definition}  \rho(x,t)=\Psi^*(x,t) \Psi(x,t).
\end{eqnarray}
These two quantities satisfy the continuity equation
\begin{eqnarray}
\label{continuity equation}  \frac{\partial \rho(x,t)}{\partial
t}+\frac{\partial J(x,t)}{\partial x}=0.
\end{eqnarray}

\begin{figure}[]
\begin{center}
\scalebox{0.8}[0.8]{\includegraphics*{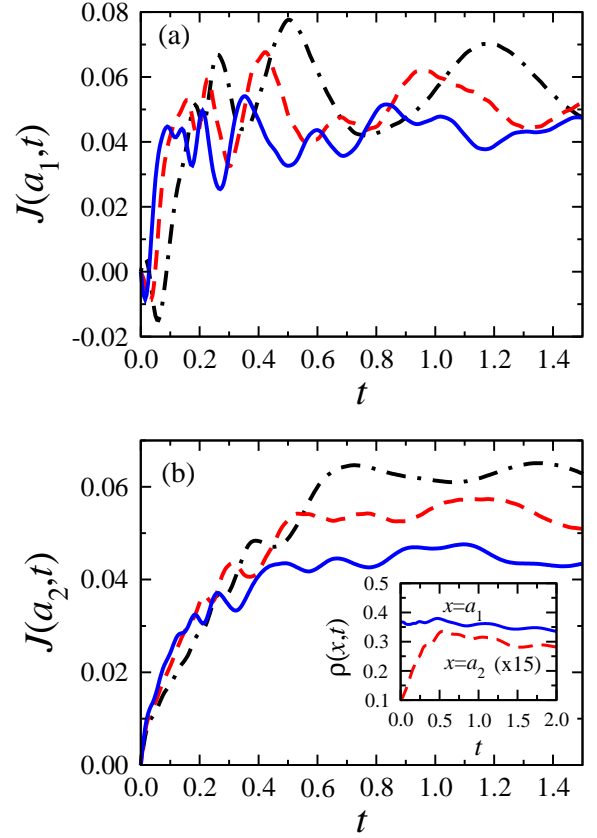}}
\caption{ (Color online)~ (a) Flux at the left edge $a_1$. The solid, dashed and dot-dashed lines
 correspond to the barriers $U_0=24$, $U_0=16$, and $U_0=10$, respectively.
 (b) Flux at the right edge $a_2$.  The inset presents the density at $a_1$
(solid) and multiplied by fifteen at $a_2$ (dashed), with the
barrier $U_0=16$. All lines correspond to the same parameters as those in (a).
For both plots, $\kappa d=2$ and $\Psi(x,0)=\varphi_0(x)$.} \label{fluxedgegnd}
\end{center}
\end{figure}
Based on Eq. (\ref{continuity equation}), we obtain the flux at the edges,
\begin{equation}
\label{flux at right edge} J(a_1,t)=-\frac{d w_1(t)}{d t},\quad J(a_2,t)=\frac{d w_2(t)}{d t},
\end{equation}
where
\begin{eqnarray} \label{probability outside}
w_2(t)=\int_{a_2}^{\infty} \Psi^* (x,t) \Psi(x,t) dx
\end{eqnarray}
is the probability to find the particle outside the potential.

\begin{figure}[]
\begin{center}
\scalebox{0.8}[0.8]{\includegraphics*{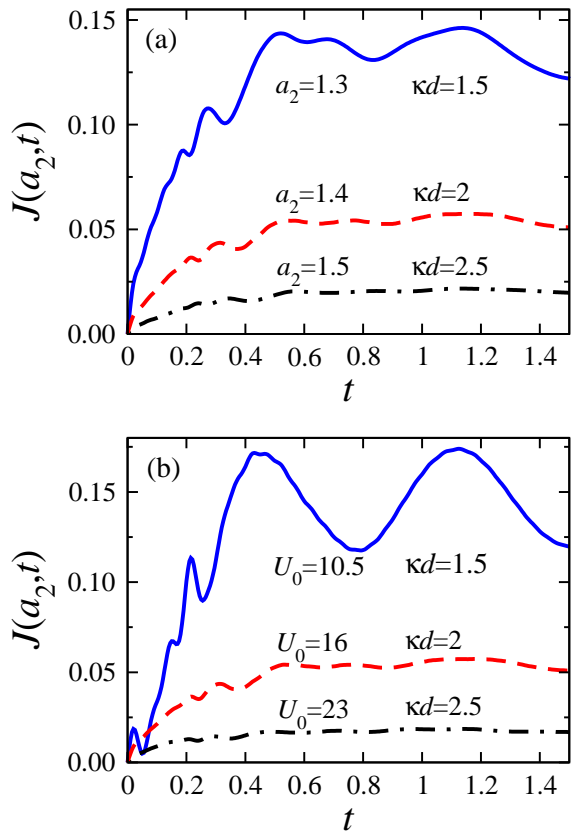}}
\caption{(Color online)~ (a) Flux at the right edge on time with different
width: solid line, $\kappa d=1.5$, dashed line, $\kappa d=2$, and dot-dashed line, $\kappa d=2.5$, $U_0=16$.
(b) Flux at the right edge on time with different height, by remaining $a_2=1.4$: solid line, $\kappa d=1.5$,
dashed line $\kappa d=2$, and dot-dashed line, $\kappa d=2.5$, respectively. The ground state energy
$E_0$ depends only weakly on the potential height $U_0.$ For all plots
$\Psi(x,0)=\varphi_0(x)$.} \label{flux d U0}
\end{center}
\end{figure}

In Fig. \ref{fluxedgegnd} we illustrate the time dependence of the
edge flux for different barriers
during a short time scale for $\Psi(x,0)=\varphi_0(x)$,
where the lines shift to the right by lowering the
barrier. It can be seen from both panels of Fig. \ref{fluxedgegnd} that
$J(a_1,t)$ and $J(a_2,t)$, basically, increase during a short time interval and then
reach at a characteristic time $t_{\rm pl}$  a rough plateau, which shows a smoother behavior for $J(a_2,t)$,
while $J(a_1,t)$ dives for a very short time to a
negative value and oscillates more strongly. As a result,
no feature can be clearly distinguished as a precise
instant when tunneling begins.
By using the continuity equations (\ref{flux at right edge}),
the decay time $t_l$ can be reliably estimated as $1/J_{\rm pl}(a_{2})$,
where $J_{\rm pl}(a_{2})$ is the typical value of the flux at the plateau.
The time scale when $J(a_2,t)$ develops a plateau also becomes larger,
although it cannot be defined precisely.
A crude estimate for the  scale at which the plateau forms is $t_{\rm pl}\sim\pi/(2E_0)$,
approximately a factor of 2 less than the oscillation
period $T_{k_0}$ of a particle in the initial well with potential $U_1$. The
period $T_{k_0}$ determines the prefactor of the escape time as discussed below
Eq. (\ref{probability1}). Therefore, classically speaking, the decay by tunneling as a steady process begins
when the electron hits the barrier.

The inset in Fig. \ref{fluxedgegnd}(b) shows the
behavior of the density at the edge. Contrary to $J(a_1,0)=J(a_2,0)=0$, $\rho(a_1,0)$ and
$\rho(a_2,0)$ are not zero, as the ground state for the potential
$U_1(x)$ is not fully localized in the well, and the plateau in $\rho(a_2,0)$
is clearly seen.

Even though the numerically estimated times are much shorter,
a similar  trend  in variation with the height of the barrier
is shown by the  B\"{u}ttiker-Landauer time (BL time), provided that $\kappa d$
remains unchanged. The traversal time of B\"{u}ttiker and Landauer
 \cite{Buttiker-Landauer}, $t_{\rm BL}=d/{\kappa}$, given
by the barrier width $d$ divided by the "semiclassical" velocity
$\kappa=[2(U_0-E_0)]^{1/2}$, is an important time scale, especially in
opaque conditions.  With the conservation of $\kappa d$, $t_{BL}$ is
proportional to $1/{\kappa^2}$.

The dependence of the flux at the right edge on time with different
width but keeping the height of the barrier is shown in
Fig. \ref{flux d U0}(a). It is interesting
to find that the time that the flux begins to form a
plateau is almost equal for the different widths,
although the flux value at the plateau changes strongly, roughly
as $\exp(-2\kappa d)$. In Fig. \ref{flux d U0}(b), this characteristic time $t_{\rm pl}$ also remains
almost unchanged for different $U_0$ by keeping the same width $a_2$.
We conclude that the observed scale of $\pi/(2E_0)$ is universal and does not depend on the
details of the potential.

For comparison, we illustrate in Fig. 6 the flux at a short time scale
for a transparent barrier $\kappa d = 0.25$. As a matter of fact, the flux and
the density have similar profiles. They grow from initial to the
maximum value and decrease rapidly, contrary to the opaque behavior
without an obvious peak. The peak is positioned at $t\approx 0.4$, similar to
the time of plateau development in Fig.\ref{flux d U0}.

\begin{figure}[] \begin{center}
 \scalebox{0.8}[0.8]{\includegraphics*{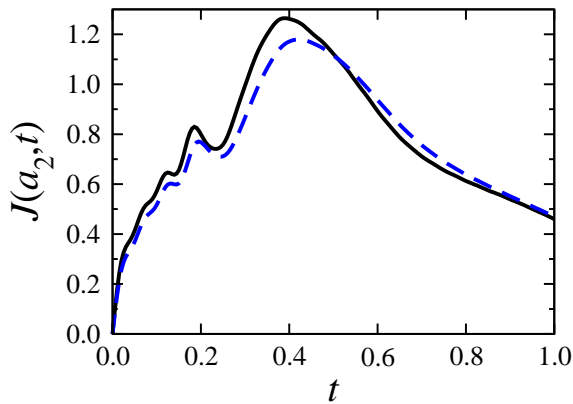}}
\caption{(Color online) Time evolution of the flux at the right edge, where the solid line represents the flux,
when $U_0=16$ and $a_2=1.05$. The dashed line corresponds to the flux for
$U_0=10$ and $a_2=1.06$. The two barriers are transparent as
$\kappa d=0.25$, and the initial state is $\varphi_0(x)$.}
\label{transparent}
\end{center}
\end{figure}

\begin{figure}[]
\begin{center}
\scalebox{0.8}[0.8]{\includegraphics*{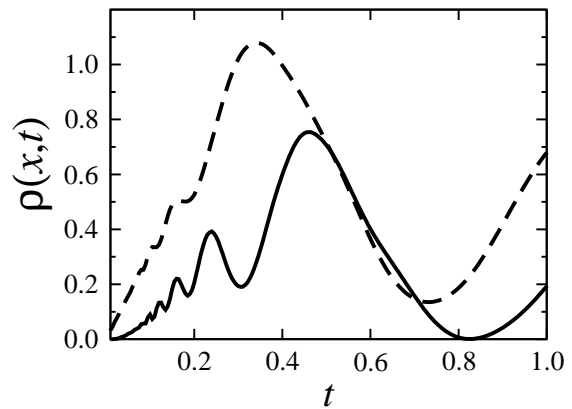}}
 \caption{Density evolution with the barriers $U_0=10$, $a_2=1.63$
 (solid line at $a_2$ and dashed line at $a_1$).
 As the density at $a_2$ is relatively small, it is
multiplied by the factor of ten. The barrier is opaque with
$\kappa d=2$ and the initial state is $\Psi(x,0)=\sqrt{2} \sin(\pi
x)$.} \label{fluxedgesin} \end{center}
\end{figure}

For the initial state $\Psi(x,0)=\sqrt{2} \sin(\pi x)$  the flux is
enhanced by an order of magnitude and oscillates stronger
compared to the initial $\Psi(x,0)=\varphi_0(x)$, because
$\sin(\pi x)$ contains larger contributions from different
eigenstates. Different from that with $\Psi(x,0)=\varphi_0(x)$,
the density is zero at $t=0$ outside the
barrier. The time-dependence of the
density at the edges with the initial state
$\Psi(x,0)=\sqrt{2} \sin(\pi x)$ is presented in Fig. \ref{fluxedgesin} for a typical barrier.
For various system parameters, the
delay time between the maximum of the density at the right and the
left edges is in a good agreement with the B\"{u}ttiker-Landauer time $t_{\rm BL}$ for these potentials.
We conclude that $t_{\rm BL}$ manifests itself as a delay time
if over-the-barrier motion becomes essential due to the choice
of the initial state. The momentum components that matter at first are the ones larger than $\sqrt{2
U_0}$, as the momentum distribution of $\sin(\pi x)$ is
broad. Moreover, its average local velocity at the right edge of the barrier
$v_{B}(a_2,t)=J(a_2,t)/\rho(a_2,t)$ decreases from a large value with some
oscillations to a relatively stable one close to $\sqrt{2 E_0}$
during a short interval, after which the real tunneling
starts. This means that the forerunners just go above
the barrier instead of tunneling. By contrast,
the tunneling process for $\Psi(x,0)=\varphi_0(x)$ occurs indeed
from the instant that the decay initiates, because $v_{B}(a_2,t)$ is always
smaller than $\sqrt{2 U_0}$.

As discussed in  \cite{MB,Del} with analytical models, $t_{BL}$ is a characteristic
time describing different phenomena, among them over-the-barrier transients.
This is rather paradoxical, considering its association with ``tunnelling''
in the defining formula, and has surely not been fully appreciated.
A more intuitive understanding of this unexpected role is still missing.

\subsection{Long-term dynamics}
We have now established that the short time dynamics of the probability flux
is governed by robust time scales. We next investigate the long term dynamics
with the goal to find out whether a suitable extrapolation of the long term
scattering data can be used to gain information on the short time dynamics.
We find that after formation of the steady tunneling, the particle probability density
shows two distinct features. The first one is an almost uniform change
in $\rho(x,t)$ for $x<a_{1}$ with $\rho(x,t)\approx\rho(x,0)\exp(-t/t_l)$.
The second one can be viewed as a broad bump (wave packet) with relatively small density
propagating with the velocity close to $\sqrt{2E_0}$ and spreading in time.
In this subsection, we consider the dynamics of the bump
at long time scale and trace it to short times to obtain the operationally defined
tunneling times.

\begin{figure}[] \begin{center}
 \scalebox{0.8}[0.8]{\includegraphics*{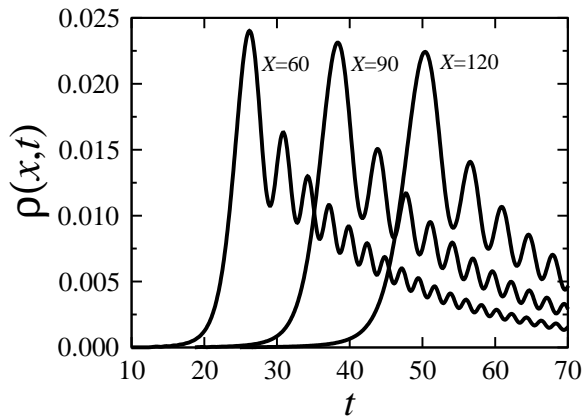}}
 \caption{Time evolution of density $\rho(X,t)$ at given positions
$X=60,90,120$, provided that $\Psi(x,0)=\varphi_0(x)$, $U_0=16$,
$a_2=1.4$. A strong initial peak is followed by oscillations due to the diffraction in time process.
} \label{density at 3 distances}
\end{center}
\end{figure}

We assume that the flux $J(X,t)$ and the density $\rho(X,t)$ are measured by a detector
located at $X\gg a_{2}$.  It is shown in Fig.
\ref{density at 3 distances} for $\Psi(x,0)=\varphi_0(x)$, that the density at $X\gg a_2$
is nearly zero up to some time, then grows to a sharp maximum, and then
decreases at timescales of $t_l$ with the sequential
oscillations due to the diffraction in time phenomenon  \cite{DIT}. The profiles of the flux and density are very similar,
with $J(X,t)\approx \sqrt{2E_0} \rho(X,t)$.

In the attosecond experiment analysis  \cite{science}
the tunneling time was defined as the delay between the time when the
barrier begins to be lowered and the time when electron experiences
the acceleration by external field, as can be extracted
from the long-time behavior. Similarly, we can use the
operational phenomenological approach to define a tunneling time
as
\begin{eqnarray}
\label{tunneling time1}  t_{\rm tun,1}=t_X-\frac{X-a_2}{\widetilde{v}_p},
\end{eqnarray}
where $t_X$ is the time when
the detector measures the strongest first peak of the flux,
$\widetilde{v}_p=\sqrt{2 \widetilde{E}_0}$ is the velocity
with which the particle moves out
of the potential, and the energy
\begin{equation}
\widetilde{E}_{0}=E_0-\int_{a_2}^\infty U_0 \varphi^2_0(x)dx
\end{equation}
is slightly less than $E_0$, as the
potential changes suddenly from $U_1(x)$ to $U_2(x)$. As $t_X$ is
a measurable quantity the tunneling time can be calculated by Eq. (\ref{tunneling time1}). For instance, for $\Psi(x,t=0)=\varphi_0(x)$, $U_0=16$,
$a_2=1.4$, when the detector is at $X=120$, then $t_X=50.3$, and resulting $t_{\rm tun,1}=5.62$.
Another approach, similar to Eq. (\ref{tunneling time1}), is to calculate the tunneling time as

\begin{eqnarray}
\label{tunneling time2}  t_{\rm tun,2}=t_X-\frac{X-a_2}{v_X},
\end{eqnarray}
where $v_X$ is the velocity of the flux peak. For example,
$v_X=2.517$ as defined by the motion from $X=120$ to $X=122$. According to
Eq. (\ref{tunneling time2}), the corresponding
tunneling time is $t_{\rm tun,2}=3.18$. As it can be assumed that
the velocity of the electron is constant outside the barrier in a
classical manner, we can conclude that the tunneling time extracted
from the measurements by a remote detector is
$X$-independent. Both Eq. (\ref{tunneling time1}) and
Eq. (\ref{tunneling time2}) extrapolate electron motion from distance
to the exit of the tunneling process. However, these two
times are not equal either to the time of formation of the steady
tunneling in Fig. \ref{fluxedgegnd} or to the decay time $t_l$.

\begin{figure}[] \begin{center}
\scalebox{0.5}[0.5]{\includegraphics{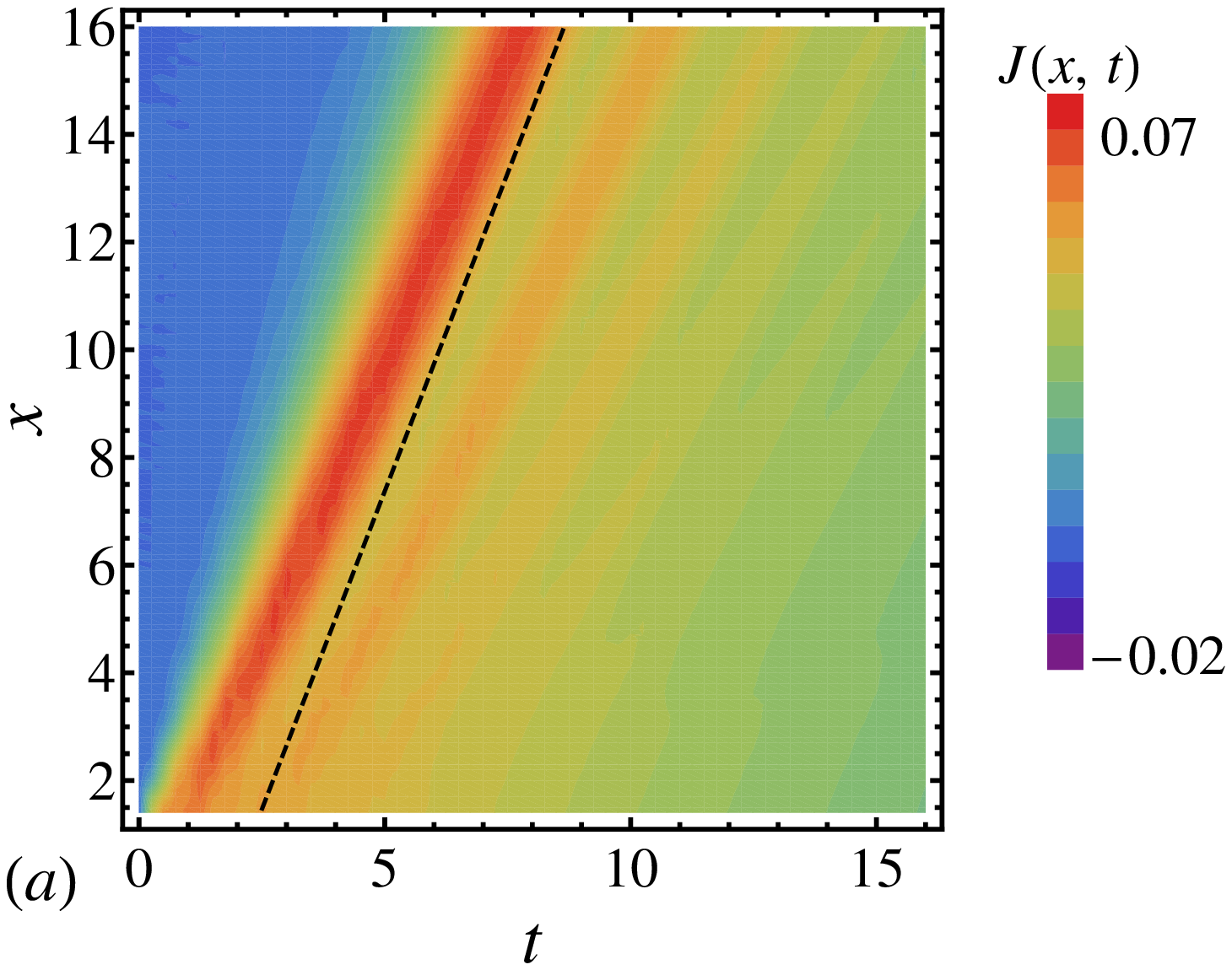}}
\scalebox{0.5}[0.5]{\includegraphics{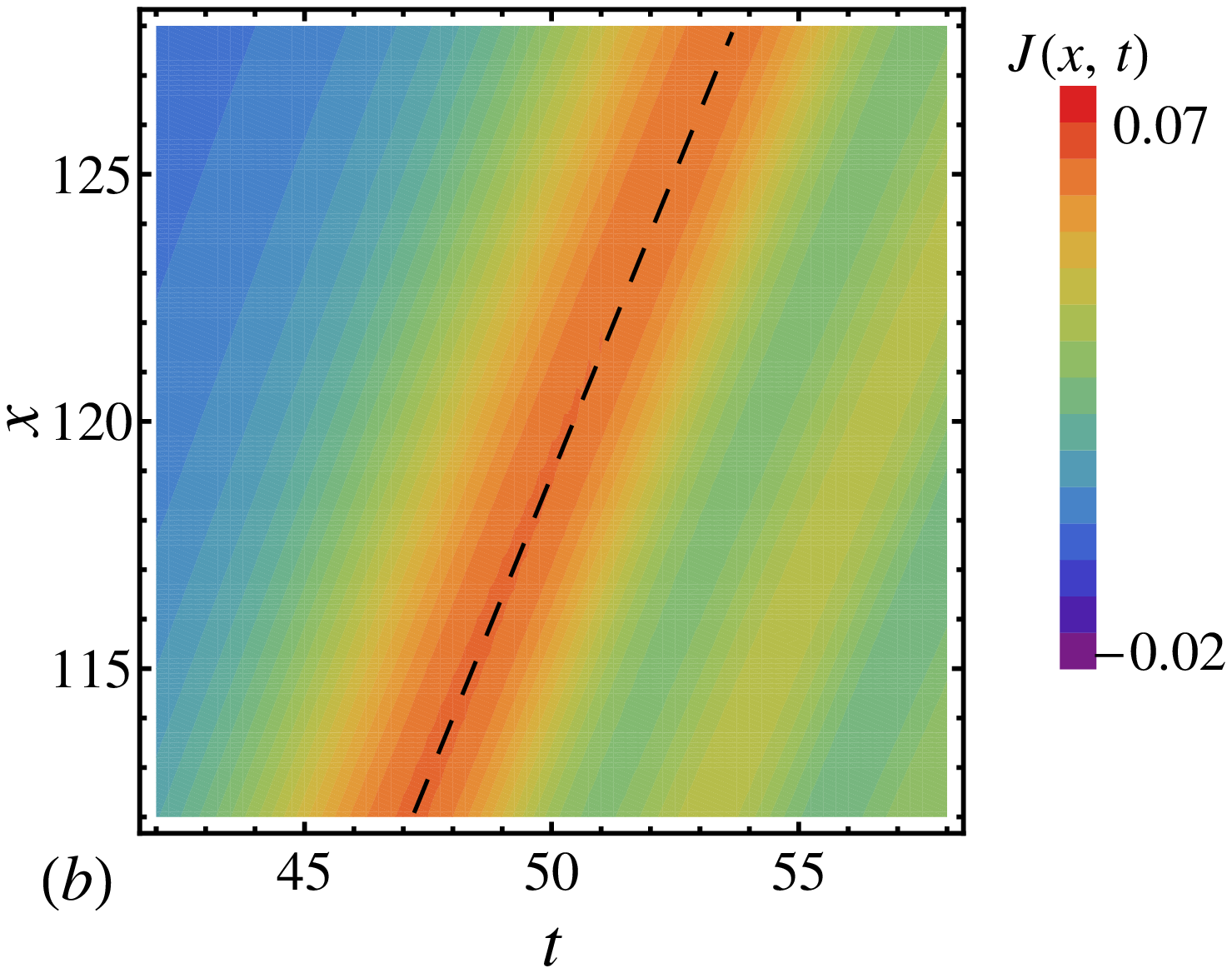}}
\caption{
(Color online) The flux versus time and position
after a short time near the barrier (a) and
after a long time far away from the barrier (b).
All the parameters are the same as in Fig. \ref{density at 3
distances}. The dashed straight lines serve as the guide for the eye only.
The short-dash line in (a) demonstrates
the nonlinearity of the peak position as a function of time. The long-dash line in (b)
shows that this dependence is very close to linear at these times
and distances.} \label{contourline}
\end{center}
\end{figure}

Fig. \ref{contourline} depicts how the flux evolves with time
and distance at two timescales for $\Psi(x,0)=\varphi_0(x)$. The curvature of the maximum
flux region in Fig. \ref{contourline}(a) shows that it takes some
time for the flux to develop a constant speed in free space.
In addition, Fig. \ref{contourline}(a) demonstrates that a peak of the flux at $a_{2}$
appears at $t\approx 1$, smaller than $t_{\rm tun,2}$ in Eq. (\ref{tunneling time2}).

It is expected in some models that the wave function at long times and distances
can be obtained with the stationary phase approximation. Eq. (\ref{wave function1}) in this limit
can be recast as
\begin{eqnarray}
\label{wave function approximation}
&&\Psi(x>a_2,t) \approx  \\ \nonumber
&&\frac{1}{\sqrt{2 \pi}i}  \int_0^\infty dk\,G(k) \exp \left\{i
\left[\theta(k)+k x-\frac{k^2 t}{2} \right ] \right\},
\end{eqnarray}
and the phase $\theta(k)+k x-k^2 t/2$ can be expanded near
the stationary point $K$ satisfying the equation $(d\theta(k)/dk)_{k=K}+x-Kt=0$.
However, $G(k)$ in our calculations is not a sufficiently smooth function near the $K$-points due to
the resonances shown in Fig. \ref{Gk}. Therefore, the stationary phase
approximation does not provide a satisfactory description of the
peak propagation.

\begin{figure}[] \begin{center}
\scalebox{0.8}[0.8]{\includegraphics*{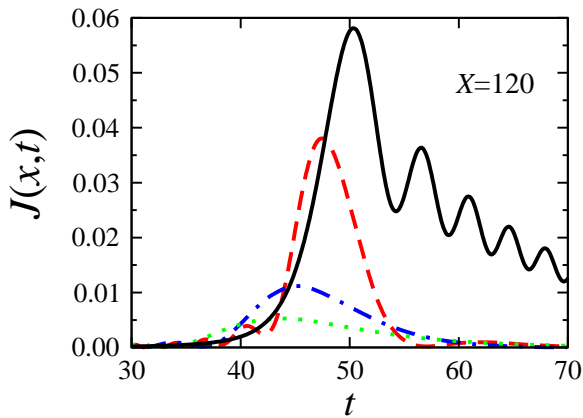}}
\caption{(Color online)~ Flux as a function of time detected
 at $X=120$ with different closing time $t_0=\infty$ (solid),
$t_0= t_l/3$ (dashed), $t_0=t_l/6$ (dot-dashed), $t_0=t_l/9$
(dotted), $t_l$ is the life time for the particle in the case of $t_0=\infty$. Other
parameters are the same as those in Fig. \ref{density at 3 distances}.} \label{flux}
\end{center}
\end{figure}

Another factor that affects the tunneling time is the closing time
$t_0$, when the potential turns back from $U_2(x)$ to $U_1(x)$ and
the tunneling is interrupted. The time evolution of the flux
observed at the same remote position of detector with different
closing times is demonstrated in Fig. \ref{flux}, and $t_l$
is the exponential decay time introduced above. The peak of the flux with shorter
closing time $t_0$ appears earlier than that with longer $t_0$, as a result of the increased energy spreading  \cite{apo}.
After the closing time $t_0$, the probability to find the particle in the well remains almost
unchanged. This shows that components with larger momenta tunnel
through the barrier first, and therefore, closing the barrier can decrease
the operationally defined times $t_{\rm tun,1}$ and $t_{\rm tun,2}$. The measurement
of the total number of particles escaped from the potential
as the function of the closing time can help to find the timescale
of the flux formation $t_{\rm pl}$: with the increase of $t_0$ through this region the number of
escaped particles which is quadratic in $t_0$ for short times changes at
$t_0>t_{\rm pl}$ to a linear dependence.

\section{Conclusions}
Motivated by recent experiments on the ultrafast atom ionization by optical
fields, we have studied numerically exactly, for a rectangular
barrier, the time-resolved tunneling for short time scales and the further
propagation at long times of an initially localized particle. The
barrier we consider is opaque, however, not extremely so, to ensure a
reasonable tunneling probability. The probability density
evolution on a short time scale much less than the decay time,  depends strongly on the
initial state. Dependent on how this state is prepared,
this short-term motion can include both under-the-barrier tunneling
and over-the-barrier propagation, as seen in the evolution of the
density and flux at the barrier edges. The tunneling starts
instantly, however, some time of the order of $\pi/(2E_0)$,
where $E_0$ is the ground state energy in the initial
potential, is required to develop the outgoing
flux eventually proportional to the characteristic exponential decay rate
$1/t_l$. As expected, there is a time delay between the flux
development at the left and the right edges of the barrier. If the initial
state is the ground state of the potential at $t<0$,
the timescale of the flux development is much longer than the
B\"{u}ttiker-Landauer traversal time expected for the given barrier
parameters. However, if the initial state is more tightly localized,
the B\"{u}ttiker-Landauer  time manifests
itself as a time delay of the flux and density maxima between the
left and right edges of the barrier.

At long times we have considered the propagation
of the escaped wave packet at distances much larger than the scale of
the potential. From the operational
definition of the escape time, related to the position of the
maximum of the wave packet density, we have estimated the time the
particles escaped from the potential. This time is also much longer
than the traversal time for the given barrier. To determine the
effect of the time dependence of the barrier, we implemented escape time
windows considerably shorter than the tunneling decay time $t_l$. The increasing
importance of faster components for shorter time windows
leads to the extrapolated time estimated for the closing potentials
smaller than that for the potentials permanently open.

\section{Acknowledgement}

E.Y.S. and J.G.M. are grateful to the support of University of Basque Country UPV-EHU
(Grant GIU07/40), Basque Country Government (IT-472-10), and  Ministry of
Science and Innovation of Spain (FIS2009-12773-C02-01). M. B. is supported by the Swiss NSF, MaNEP, and the European ITN, NanoCTM.


\begin{thebibliography}{99}
\bibitem{MacColl} L. A. MacColl, Phys. Rev. \textbf{40}, 621 (1932).
\bibitem{Mugabook} \textit{Time in Quantum Mechanics}, Ed. by J. G. Muga, R. Sala Mayato, and I. L. Egusquiza (Springer, Berlin, 2002)
\bibitem{Wigner} E. P. Wigner, Phys. Rev. \textbf{98}, 145 (1955).
\bibitem{Hartman} T. E. Hartman, J. Appl. Phys. \textbf{33}, 3427 (1962).
\bibitem{Baz} A. I. Baz, Sov. J. Nucl. Phys. \textbf{4}, 182 (1967); \textbf{5}, 161 (1967).
\bibitem{Rybachenko} V. F. Rybachenko, Sov. J. Nucl. Phys. \textbf{5}, 635 (1967).
\bibitem{Sherman10}  D. V. Khomitsky and E. Ya. Sherman,
	             Europhys. Lett. {\bf 90}, 27010 (2010).
\bibitem{Appelbaum10} B. Huang and I. Appelbaum, Phys. Rev. B \textbf{82}, 241202 (2010).
\bibitem{dwell} M. B\"uttiker, Phys. Rev. B \textbf{27}, 6178  (1983).
\bibitem{Buttiker-Landauer} M. B\"{u}ttiker and R. Landauer, Phys. Rev. Lett. \textbf{49}, 1739 (1982).
\bibitem{Chiao-optical barrier} A. M. Steinberg, P. G. Kwiat, and R. Y. Chiao, Phys. Rev.
Lett. \textbf{71}, 708 (1993).
\bibitem{Chiao-photonic band} Ch. Spielmann, R. Szip\"{o}cs, A. Stingl, and F. Krausz, Phys. Rev. Lett. \textbf{73},
2308 (1994).
\bibitem{Stein} A. M. Steinberg, Lect. Notes Phys. {\bf 734}, 333 (2008).
\bibitem{Mugn} D. Mugnai and A. Ranfagni, Lect. Notes Phys. {\bf 734}, 355 (2008).
\bibitem{Sepkhanov09}  R. A. Sepkhanov, M. V. Medvedyeva, and C. W. J. Beenakker,
                       Phys. Rev. B {\bf 80}, 245433 (2009).
\bibitem{BSM} S. Brouard, R. Sala, and J. G. Muga, Phys. Rev. A {\bf 49}, 4312 (1994).
\bibitem{BM} S. Brouard and J. G. Muga, Phys. Rev. A \textbf{54}, 3055 (1996).
\bibitem{Buttiker_Thomas} M. B\"uttiker and H. Thomas, Ann. Phys. (Leipzig) {\bf 7}, 602 (1998).
\bibitem{transient} A. del Campo, G. Garc\'\i a Calder\'on, and J. G. Muga,
Phys. Rep. \textbf{476}, 1 (2009).
\bibitem{DIT} M. Moshinsky, Phys. Rev. \textbf{88}, 625 (1952).
\bibitem{density of states} V. Gasparian, T. Christen, and M. B\"{u}ttiker, Phys. Rev. A \textbf{54},
4022 (1996); M. B\"{u}ttiker, J. Phys. (Pramana) \textbf{58}, 241
(2002).
\bibitem{Decay1} A. del Campo, F. Delgado, G. Garc\'{\i}a-Calder\'{o}n, J. G. Muga, and M. G. Raizen, Phys. Rev. A \textbf{74}, 013605 (2006).
\bibitem{Decay2} A. Marchewka and E. Granot, Phys. Rev. A \textbf{79}, 012106 (2009).
\bibitem{Kalbermann} G. K\"{a}lbermann, Phys. Rev. C \textbf{79}, 024613 (2009), G. K\"{a}lbermann, Phys. Rev. C \textbf{77}, 041601
(2008).
\bibitem{Kelkar-EPL} N. G. Kelkar, H. M. Casta\~{n}eda, and M. Nowakowski, EPL \textbf{85}, 20006 (2009).
\bibitem{Buttiker93} M. B\"{u}ttiker, H. Thomas, and A. Pretre, Phys. Lett. A {\bf 180}, 364 (1993).
\bibitem{Gabelli} J. Gabelli, G. Feve, J.-M. Berroir, B. Placais, A. Cavanna, B. Etienne, Y. Jin, and D. C. Glattli,
                 Science {\bf 313}, 499 (2006).
\bibitem{Nigg06} S.E. Nigg, R. Lopez, and Markus B\"{u}ttiker,
                 Phys. Rev. Lett. {\bf 97}, 206804 (2006).
\bibitem{Fev2007} G. F\'eve, A. Mah\'e, J.-M. Berroir, T. Kontos, B. Placais,
                  D.C. Glattli, A. Cavanna, B. Etienne, and Y. Jin,
                  Science {\bf 316}, 1169 (2007).
\bibitem{Moskalets08} M. Moskalets, P. Samuelsson, and M. B\"{u}ttiker,
                     Phys. Rev. Lett. {\bf 100}, 086601 (2008).
\bibitem{Keeling08} J. Keeling, A. V. Shytov, and L. S. Levitov,
                    Phys. Rev. Lett. {\bf 101}, 196404 (2008).
\bibitem{Buttiker10} M. B\"{u}ttiker and M. Moskalets,
                    Int. Journ. Mod. Phys. B {\bf 24}, 1555 (2010).
\bibitem{Dmitri} D. Sokolovski, Lect. Notes Phys. \textbf{734}, 195 (2008).

\bibitem{science} P. Eckle, A.N. Pfeiffer, C. Cirelli, A. Staudte, R. D\"{o}ner, H.G. Muller,
                  M. B\"{u}ttiker, and U. Keller,
                  Science \textbf{322}, 1525 (2008).

\bibitem{NRC} L. Arissian, C. Smeenk, F. Turner, C. Trallero, A. V. Sokolov,
              D. M. Villeneuve, A. Staudte, and P. B. Corkum,
              Phys. Rev. Lett. \textbf{105}, 133002 (2010)

\bibitem{Kelkar1} N. G. Kelkar, M. Nowakowski, K. P. Khemchandani, and S. R. Jain,
                  Nucl. Phys. A \textbf{730} 121 (2004).
\bibitem{Kelkar2}   N. G. Kelkar,  K. P. Khemchandani,  and   B. K. Jain,
                    J. Phys. G: Nucl. Part. Phys. \textbf{32} 1157 (2006).

\bibitem{MB} J. G. Muga and M. B\"uttiker, Phys. Rev. A \textbf{62}, 023808 (2000).
\bibitem{Del}F. Delgado, J. G. Muga, A. Ruschhaupt, G. Garc\'\i a-Calder\'on,  and J. Villavicencio, Phys. Rev. A \textbf{68}, 032101 (2003).
\bibitem{apo} A. del Campo, J. G. Muga and M. Moshinsky, J. Phys. B: At. Mol. Opt. Phys. \textbf{40}, 975 (2007).


\end{thebibliography}
\end{document}